\documentclass[prb,twocolumn,showpacs,superscriptaddress]{revtex4}
\usepackage{graphicx}

\begin{document}

\title {Elastic electron scattering in quantum corrals: The importance of the shape
of the adatom potential.}
\author{A. I. Rahachou}\email{alira@itn.liu.se}
\affiliation{ Department of Science and Technology (ITN),
Link\"{o}ping University, S--601\,74 Norrk\"oping, Sweden}
\author{I. V. Zozoulenko}\email{igozo@itn.liu.se}
\affiliation{ Department of Science and Technology (ITN),
Link\"{o}ping University, S--601\,74 Norrk\"oping, Sweden}

\date{\today}

\begin{abstract}

We report elastic scattering theory for surface electron waves in
quantum corrals defined by adatoms on the surface of noble metals.
We develop a scattering-matrix technique that allows us to account
for a realistic smooth potential profile of the scattering
centers. Our calculations reproduce quantitatively all the
experimental observations, which is in contrast to previous
theories (treating the adatoms as point scatterers) that require
additional inelastic channels of scattering into the bulk in order
to achieve the agreement with the experiment. Our findings thus
indicate that surface states are not coupled to the bulk
electrons.

\end{abstract}
\pacs{72.10.Fk, 73.20.At, 68.73.Ef, 03.65.Nk}

\maketitle

Advances in modern nanotechnology made it possible to manipulate
adatoms on the surface of a metal, arranging them into ordered
structures coined as ``quantum corrals". Using the scanning
tunneling microscopy (STM), Crommie \textit{et
al.}\cite{Crommie_Nature, Crommie_Science, Heller_RMP} studied the
scattering of the surface electron waves residing at (111) faces
of Cu. These surface states interact strongly with Fe adatoms, and
the spatial variation of the STM differential conductance revealed
beautiful images of the standing wave patterns in the quantum
corrals. In addition, the experiment showed a series of pronounced
peaks in the energy spectrum of the differential conductance
$dI/dV$ at the center of the corrals.

In order to describe the experimental
observation\cite{Crommie_Nature, Crommie_Science}, Heller
\textit{et al.}\cite{Heller} have developed the
multiple-scattering theory for surface electron waves in quantum
corrals. In their theory each adatom was modelled as a point-like
$\delta$-function potential supporting isotropic $s$-wave
scattering. The quantitative agreement with the experiment was
achieved by assuming an additional (inelastic) channel of
scattering (presumably into the bulk metallic states). The authors
concluded that absorption is the dominant mechanism for the
broadening of the energy levels seen in the experiment, and
estimated that $\sim 25\%$ of the incident amplitude is reflected,
$\sim 25\%$ is transmitted, and $\sim 50\%$ is absorbed. The
importance of the electron scattering to the bulk states for the
level width broadening in the quantum corrals was also asserted by
Crampin \textit{et al.}\cite{Inglesfield} and Cramplin and
Bryant\cite{Crampin}.

An alternative purely elastic scattering theory for the same
quantum corral structures was reported by Harbury and
Porod\cite{Porod}. They modelled the adatoms by finite-height
potential barriers, as opposed to ``black dot" $\delta$-point
absorbing scattering potential adopted in the above cited
works\cite{Heller,Inglesfield,Crampin}. The elastic theory
accounts well for the spatial variation of the wave function in
the quantum corrals, but overestimates the broadening of the
resonant levels, especially for higher energies. The findings of
Harbury and Porod therefore suggest that the features of the
spectrum can be sensitive to the detailed shape of the scattering
potential.

It is important to stress that accounting for a detailed shape of
a scattering potential was crucial for quantitative description of
many phenomena in quantum nanostructures. This for example
includes the Hall and bend resistance anomalies in four-terminal
junctions\cite{Baranger}, the breakdown of quantized conductance
in quantum point contacts calculated using realistic
potentials\cite{Davies}, the explanation of a branched flow in a
two-dimensional electron gas\cite{Topinka}. In the present Brief
report we develop a scattering matrix approach that allows us to
account for a realistic smooth potential of the adatoms.
%We show
%that calculated broadening and positions of the resonant states of
%the quantum corrals as well as the scattering wave function agrees
%quantitatively with the experimental data without any additional
%(inelastic) scattering channels.
We demonstrate that for such a potential the broadening and
positions of the resonant states as well as the scattering wave
function in the quantum corrals can be quantitatively described by
the inelastic theory alone, without the assumption of any
additional (inelastic) scattering channels. Our findings thus
support the conclusions of Harbury and Porod \cite{Porod} that the
adatoms do not significantly couple the surface-state electrons to
the bulk metallic states.

The differential conductance $dI/dV$ of the STM tunnel junction is
proportional to the local density of states (LDOS) which is given
in terms of the scattering eigenstates of the Hamiltonian
$\hat{H}$, $\psi_{q}(\mathbf{r})$\cite{Heller_RMP},
\begin{eqnarray}
dI/dV\sim
\textrm{LDOS}(\mathbf{r},E)=\sum_q|\psi_{q}(\mathbf{r})|^2\delta({E-E_q}).
\end{eqnarray}

In order to calculate the scattering eigenstates
$\psi_{q}(\mathbf{r})$ we adopt to the problem at hand the
scattering matrix technique\cite{Rahachou} that was recently
developed for numerical solution of the Helmholtz equation for
resonant states of dielectric optical cavities with both complex
geometry and variable refraction index. This is possible because
of a direct correspondence between the Helmholtz and Schr\"odinger
equations\cite{Datta}. The advantage of the scattering matrix
technique is that it provides an efficient way to treat the smooth
realistic profile of the adatom. Note that commonly used methods
based on the discretization of the scattering domain would be
rather impractical in terms of both computation power and memory,
because the smoothly varying potential of the adatom has to be
mapped into a discrete grid with a very small lattice constant.

We consider a two-dimensional ring-shaped corral structure.
Experimental observations suggest that adatoms strongly disturb
the local charge density at the finite distance $\sim 7$\AA
\cite{Heller_RMP}. An exact experimental shape of the adatom
potential is not available. We thus model this potential as a
Gaussian with the half-width $\sigma$ and the height $V_0$
centered at the location $(x_0,y_o)$, $V(x,y) = V_0
\exp[{-(x-x_0)^2}/{2\sigma^2}] \exp[{-(y-y_0)^2}/{2\sigma^2}]$,
(see below, inset to Fig. \ref{fig3}).

In order to calculate the scattering eigenstates
 we divide the quantum corral in inner,
outer and intermediate regions. In the inner region (inside the
corral) and in the outer region (outside the corral) the adatom
potential is negligible, $V(x,y)=0$. Therefore, in these two
regions the solution to the Schr\"odinger equation can be written
in analytical form.
%Introducing the polar coordinates we can write
%for the wave function inside the corral,
%\begin{equation}\label{psi_in}
% \Psi^{in}=\sum_{q=-\infty}^{+\infty}a_q J_q(kr)
% e^{iq\varphi},
% \end{equation}
%where $J_q$ is the Bessel function of the order $q$, and
%$k=\sqrt{2m^*E}/\hbar$, with $m^*$ being the effective electron
%mass. The wave function outside the corral reads,
Introducing the polar coordinates we can write for the wave
function outside the corral
 \begin{equation}\label{psi_out}
 \Psi^{out}=\sum_{q=-\infty}^{+\infty}\left(A_q H_q^{(2)}(kr)+B_q H_q^{(1)}(kr)
 \right)e^{iq\varphi},
 \end{equation}
where $H_q^{(1)},H_q^{(2)}$ are the Hankel functions of the first
and second kind of the order $q$ describing respectively incoming
and outgoing waves; $k=\sqrt{2m^*E}/\hbar$, with $m^*$ being the
effective electron mass. The expression for the wave function
inside the corral $\Psi^{in}$ can be written in a similar fashion
as an expansion over Bessel functions $J_q$.

We introduce the scattering matrix $\mathbf{S}$ in a standard
fashion\cite{Datta,Rahachou}, $B=\mathbf{S}A$,
%\begin{equation}\label{scatter_matrix}
%B=\mathbf{S}A,
% \end{equation}
where $A, B$ are column vectors composed of the expansion
coefficients $A_q,B_q$ for incoming and outgoing states in Eq.
(\ref{psi_out}). The matrix element $\mathbf{S}_{q'q}$ gives the
probability amplitude of scattering from an incoming state $q$
into an outgoing state $q'$. In order to apply the scattering
matrix technique we divide the intermediate region (i.e. the
region where the adatom potential $V(x,y)$ is distinct from zero)
into $N$ narrow concentric rings. At each $i$-th boundary between
the rings we introduce the scattering matrix $\mathbf{S^i}$ that
relates the states propagating (or decaying) towards the boundary,
with those propagating (or decaying) away from the boundary. The
matrices $\mathbf{S^i}$ are derived using the requirement of
continuity of the wave function and its first derivative at the
boundary between the two neighboring rings. Successively combining
the scattering matrixes for all the
boundaries\cite{Datta,Rahachou},
$\mathbf{S^1}\otimes\ldots\otimes\mathbf{S^{N}}$, we can relate
the combined matrix to the scattering matrix $\mathbf{S}$. With
the help of the scattering matrix $\mathbf{S}$ we determine the
wave function in the outer region for every incoming state $q$ in
Eq. (\ref{psi_out}). Using the expression for the matrixes
$\mathbf{S^i}$ we then recover the corresponding wave functions in
the intermediate region as well as the wave function $\Psi^{in}$
in the inner region.

Note that in the scattering matrix technique one combines only two
scattering matrixes at each step. Hence, it is not necessary to
keep track of the solution for the wave function in the whole
space. This obviously eliminates the need for storing large
matrices and facilitates the computational speed.

\begin{figure}[!htp]

\includegraphics[scale=0.33]{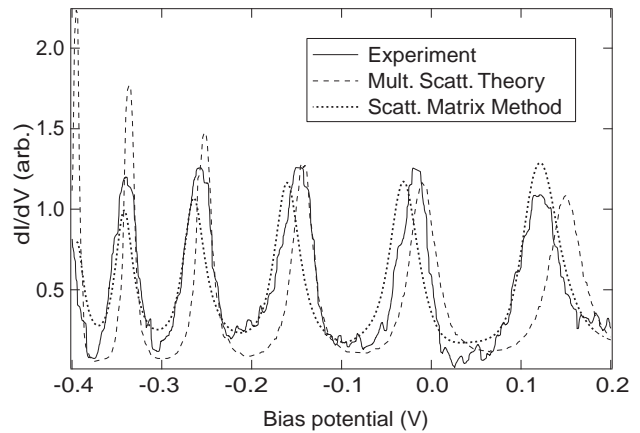}

\caption{The experimental spectrum of the differential conductance
$dI/dV$ at the center of the 88.7-\AA-radius 60-Fe-adatom circular
quantum corral structure on Cu(111) substrate (adopted from Ref.
\onlinecite{Heller}) (solid curve). The calculated spectrum,
dotted line: Our scattering matrix technique applied for a smooth
adatom potential with $V_0 = 2.5 eV$, $\sigma = 1.52 \AA$; dashed
line: Multiple-scattering theory for $\delta$-barrier adatom
potential with inelastic channel of scattering (adopted from Ref.
\onlinecite{Heller}).} \label{fig1}

\end{figure}

\begin{figure}[!htp]

\includegraphics[scale=0.8]{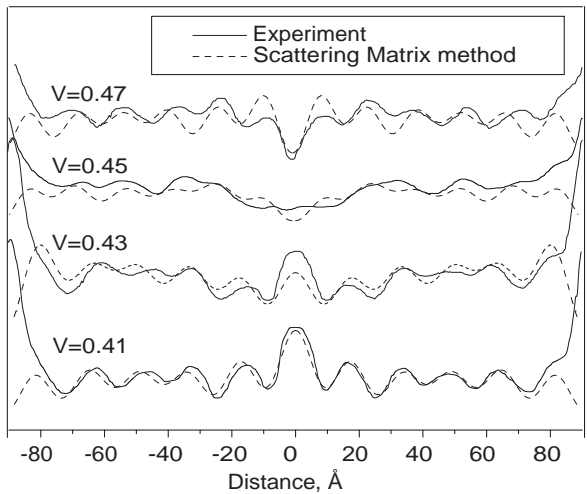}

\caption{The experimental curves (solid lines, adapted from Ref.
\onlinecite{Heller}) for the LDOS subject to the tip position
inside a circular corral for low bias voltages. The calculated
LDOS, dashed line: Our scattering matrix technique applied for a
smooth adatom potential with $V_0 = 2.5 eV$, $\sigma = 1.52 \AA$;
dot-dashed line: Multiple-scattering theory for $\delta$-barrier
adatom potential with inelastic channel of scattering (adopted
from Ref. \onlinecite{Heller}). Parameters of the structure and
are the same as those in Figs. \ref{fig1}. Voltages are given in
volts and measured relatively to the bottom of the surface-state
band. All the theory voltages are shifted by -0.01 V relatively to
the experiment.} \label{fig2}

\end{figure}

Using our scattering matrix technique we calculate the bias
voltage dependence and the spatial distribution of the LDOS for
60-Fe-adatom, 88.7-\AA-radius circular quantum corrals reported by
Heller \textit{et al.} \cite{Heller}(Figs. \ref{fig1} and
\ref{fig2}). The Fe-adatoms are placed on the meshes of a 2.55\AA\
triangular grid corresponding to the hexagonal Cu(111) lattice.
The effective mass used in all the simulations was taken $m^* =
0.361m_0$ and the electron band-edge energy $E_0=0.43$eV below the
Fermi energy of the electrons\cite{Crommie_Nature, Porod}. In the
absence of applied voltage $V$ these parameters correspond to the
wavelength of electrons $\lambda=30$\AA\ . For the parameters of
the adatom potential we use \cite{Porod} $V_0 = 2.5 eV$, $\sigma =
1.52 \AA$ which correspond to those used by Harbury and
Porod\cite{Porod} who modelled the adatoms as hard wall finite
potential barriers of 1.52-\AA\ diameter of the height of 2.5 eV.

Figures \ref{fig1} also shows corresponding results of the
multiple scattering theory of Heller \textit{et al.}\cite{Heller}.
Both theories show a similar level of agreement with the
experimental data in both the peak positions and level broadenings
for the differential conductance as well as in the number and the
peak positions for the spatial LDOS distribution throughout the
cross-section of the quantum corral. But in contrast to the case
of $\delta$-scatterers used in Ref. [\onlinecite{Heller}] our
model agrees quantitatively well with the experimental data
without introduction additional inelastic scattering channels.

%_______________________________________________________________________________
%_______________________________________________________________________________

%_______________________________________________________________________________
%_______________________________________________________________________________

Figure \ref{fig3} represents the differential conductance spectra
$dI/dV$ at the center of the quantum corral structure for various
widths of adatom potentials $\sigma$. The increase in the
potential width leads to significant narrowing of the peaks in the
spectra due to stronger confinement of the wave function inside
the corral. The best agreement with the experiment is achieved for
$\sigma\approx 1.5\AA$, which is in agrement with the results
reported by Harbury and Porod\cite{Porod}. It is often assumed
that because the spatial extent of the scattering potential
($\sigma \sim 7\AA$) is small compared to the wavelength of the
incoming particles ($\lambda\sim 30\AA$), the adatom potential can
be treated as a point scatterer or even as a continuous boundary
$V(r)\sim \delta (r-r_0)$\cite{Heller,Heller_RMP,Lobos}. The
results presented in Fig. \ref{fig3} clearly show that even though
$\sigma\ll\lambda$, the finite width of the scattering potential
affects strongly the observed characteristic of the systems. Our
calculations thus signify the importance of the shape of the
scattering potential for achieving the quantitative agreement with
the experiment of the voltage dependence of $dI/dV$.

%_______________________________________________________________________________
\begin{figure}[!htp]

\includegraphics[scale=0.33]{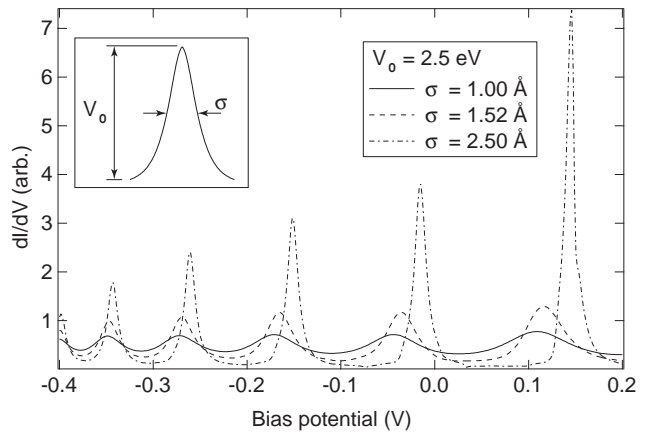}

\caption{The calculated spectrum of the differential conductance
$dI/dV$ at the center of the circular quantum corral structure for
various widths $\sigma$ of adatom potential for fixed barrier
height $V_0 = 2.5 eV$. Parameters of the structure are the same as
those in Fig. \ref{fig1}. Inset shows the schematic shape of the
potential.} \label{fig3}

\end{figure}
%_______________________________________________________________________________

%_______________________________________________________________________________
\begin{figure}[!htp]

\includegraphics[scale=0.33]{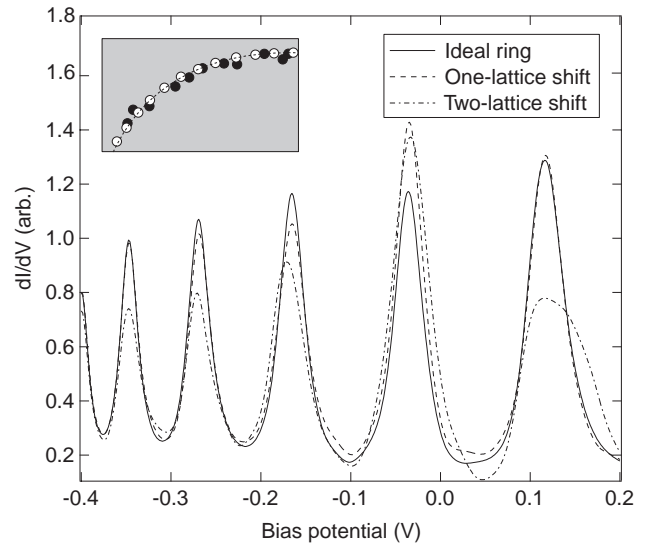}

\caption{Effect of the non-ideal positioning of the adatoms on the
differential conductance of the quantum corral. Parameters of the
structure and the adatom potential are the same as those in Figs.
\ref{fig1} and \ref{fig2}. Inset illustrates the displacement of
the scatterers from their ideal positions (white circles) in a
circular geometry for two-lattice shift.} \label{fig4}

\end{figure}
%_______________________________________________________________________________

The experiment\cite{Crommie_Science} suggests that for high
energies of incoming electrons (i.e. for large tip voltages) the
significant fraction of Fe ad atoms can move from their original
positions. We therefore study the effect of this displacement on
the shape and broadening of the resonant states of the quantum
corral. Figure \ref{fig4} shows the calculated differential
conductance for a quantum corral for the case when the adatoms of
an ideal circular corral are randomly shifted from their positions
by one and two lattice constant (as illustrated in the insets to
Fig. \ref{fig4}). One-lattice shift does not seem to have a
significant effect on the broadening of the peaks in the
differential conductance. The deviation from an ideal circular
case become rather noticeable for the shift in adatom positions by
two lattice sites. As expected, these deviations are more
pronounced for larger energies of incoming electrons. Note that we
performed simulations for different realization of the ensembles
of scatterers (keeping the average shift fixed to one or two
lattice sites), and all of them are almost indistinguishable. This
is because of the self-averaging character of scattering in a
quantum corral due to a large number of scatterers. Our
calculations thus pinpoint the deviation from a regular
arrangements of scatterers as an additional factor that should be
taken into account for achieving the quantitative agreement with
the experiment.

To conclude, we studied the scattering of electron waves by
quantum corral structure for the case of realistic smooth
potential of adatom scatterers. We achieved a detailed agreement
with the experiment without introducing the additional inelastic
channels of scattering. Our results thus support the theory that
adatoms do not significantly couple the surface-state electrons to
the bulk states. Our findings also suggest that accounting for a
realistic potential shape may be of particular importance for the
quantitative description of quantum mirages experiments.

%\begin{acknowledgments}
Financial support from Vetenskapsr\aa det (I.V.Z) and the National
Graduate School of Scientific Computing (A. I. R.) is gratefully
acknowledged. We appreciate a discussion with G. Hansson.
%\end{acknowledgments}

\end{document}